\documentclass[10pt,prl,twocolumn,superscriptaddress]{revtex4-1}

\usepackage[dvips]{graphicx} % for figures
\usepackage{amsfonts}
\usepackage{amssymb}
\usepackage{amscd}
\usepackage{amsmath}    % need for subequations
\usepackage{enumerate}
\usepackage{epsfig}
\usepackage{subfigure}
\usepackage{xcolor}
\usepackage{amsthm}

\newcommand{\bra}[1]{\mbox{$\left\langle #1 \right|$}}
\newcommand{\ket}[1]{\mbox{$\left| #1 \right\rangle$}}

\newcommand{\rhoin}{\rho^{\mathrm{(in)}}}
\newcommand{\rhoout}{\rho^{\mathrm{(out)}}}
\newcommand{\rhoctc}{\rho_{\mathrm{CTC}}}

\begin{document}
%\title{Universal Quantum Cloning by Erasing Quantum Correlations}

\title{Replicating the benefits of closed timelike curves without breaking causality}
\
\date{\today}% It is always \today, today,
\author{Xiao Yuan}
\affiliation{Center for Quantum Information, Institute for Interdisciplinary Information Sciences, Tsinghua University, Beijing, China}

\author{Syed M. Assad}
\affiliation{Centre for Quantum Computation and Communication Technology, Department of Quantum Science,
The Australian National University, Canberra, ACT 0200, Australia}

\author{Jayne Thompson}
\affiliation{Centre for Quantum Technologies, National University of Singapore, 3 Science Drive 2, 117543 Singapore, Singapore}

\author{Jing Yan Haw}
\affiliation{Centre for Quantum Computation and Communication Technology, Department of Quantum Science, The Australian National University, Canberra, ACT 0200, Australia}

\author{Vlatko Vedral}
\affiliation{Clarendon Laboratories, Oxford University, United Kingdom}
\affiliation{Centre for Quantum Technologies, National University of Singapore, 3 Science Drive 2, 117543 Singapore, Singapore}

\author{Timothy C. Ralph}
\affiliation{Centre for Quantum Computation and Communication Technology School of Mathematics and Physics, University of Queensland, St Lucia, Queensland 4072, Australia}

\author{Ping Koy Lam}
\affiliation{Centre for Quantum Computation and Communication Technology, Department of Quantum Science, The Australian National University, Canberra, ACT 0200, Australia}

\author{Christian Weedbrook}
%\affiliation{Department of Physics, University of Toronto, Toronto, M5S 3G4, Canada}
\affiliation{QKD Corp., 60 St. George St., Toronto, M5S 3G4, Canada}

\author{Mile Gu}
\email{cqtmileg@nus.edu.sg}
\affiliation{Center for Quantum Information, Institute for Interdisciplinary Information Sciences, Tsinghua University, Beijing, China}
\affiliation{Centre for Quantum Technologies, National University of Singapore, 3 Science Drive 2, 117543 Singapore, Singapore}

\begin{abstract} In general relativity, closed timelike curves can break causality with remarkable and unsettling consequences. At the classical level, they induce causal paradoxes disturbing enough to motivate conjectures that explicitly prevent their existence. At the quantum level, resolving such paradoxes induces radical benefits - from cloning unknown quantum states to solving problems intractable to quantum computers. Instinctively, one expects these benefits to vanish if causality is respected. Here we show that in harnessing entanglement, we can efficiently solve NP-complete problems and clone arbitrary quantum states - even when all time-travelling systems are completely isolated from the past. Thus, the many defining benefits of closed timelike curves can still be harnessed, even when causality is preserved. Our results unveil a subtle interplay between entanglement and general relativity, and significantly improve the potential of probing the radical effects that may exist at the interface between relativity and quantum theory.
\end{abstract}

%\pacs{}% PACS, the Physics and Astronomy
%\keywords{}%Use showkeys class option if keyword
\maketitle

Causality aligns with our natural sense of reality. We expect there to be a natural chronology to our reality -  two events should not be simultaneous causes for each other. The breaking of causality defies classical logic, resulting in causal paradoxes with no simple solution - the iconic example being the case where a man travels back in time to kill his own grandfather. Thus, physical predictions that break causality face intense scrutiny - often considered to be theoretical artifacts that are likely suppressed once we gain a more complete understanding of reality - motivating various chronology protection conjectures~\cite{Hawking92}.

Nevertheless, causality breaking theories are consistent with current scientific knowledge. Closed timelike curves (CTCs) are valid solutions of Einstein's equations in general relativity~\cite{Godel49,Morris88,Gott91}. Meanwhile, Deustch demonstrated that in the quantum regime, the resulting causal paradoxes always have self consistent solutions~\cite{deutsch1991quantum}. This resolution, however, has radical operational consequences. Many foundational constraints of quantum theory break. Non-orthogonal quantum states can be perfectly distinguished, the uncertainty principle can be violated, and arbitrary unknown quantum states can be cloned to any fixed fidelity~\cite{Brun09,brun2013quantum}. In harnessing these effects, many problems thought to be intractable to standard quantum computers now field efficient solutions~\cite{brun2003computers,Bacon04,aaronson2009closed,aaronson2005guest}. Though radical, these effects seem somewhat rationalized in the context of requiring broken causality - the sentiment being that they are curiosities that will vanish when causality is imposed.

\begin{figure}
\begin{center}
\includegraphics[width=0.4\textwidth]{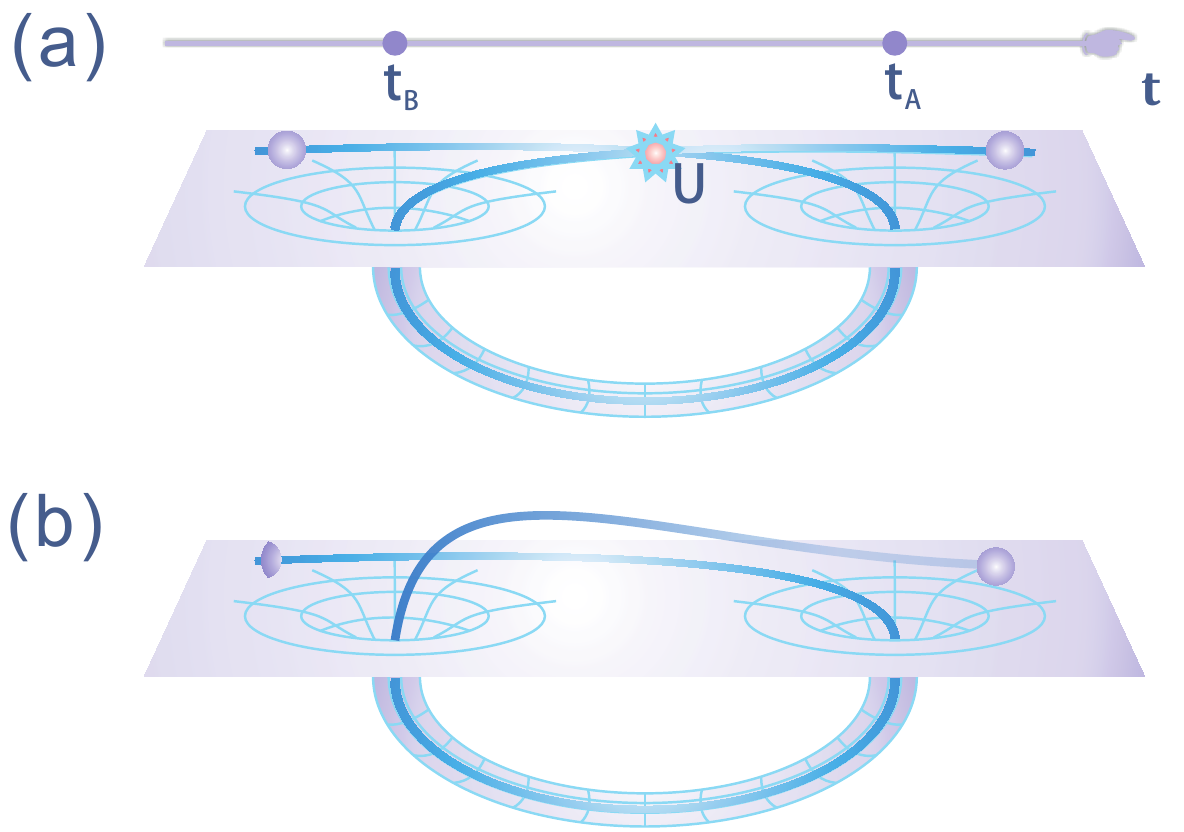}
\end{center}
 \caption{\textbf{Deutschian timelike curves}. (a) depicts a physical visualization of a CTC, where an object entering one mouth of a wormhole at some point $t_A$ may jump to a prior time $t_B$ (with respect to an chronology respecting observer) and interact with its past self via some unitary $U$. (b) In the special case where no interaction occurs, we obtain an open timelike curve. This naturally occurs, for example, in instances where the wormhole mouth are spatially separated.}\label{Fig:CTC}
\end{figure}

What happens, however, if causality is not strictly broken? In this context, Pienaar et.al introduced open timelike curves~\cite{Pienaar13} (OTCs). Consider a particle that travels back in time with respect to a chronology respecting observer, but is completely isolated from anything that can affect its own causal past during the time-traveling process (See Fig. \ref{Fig:CTC}). While the time-traveling particle has the potential to break causality, its complete isolation ensures that causality never breaks. Nevertheless, such OTCs can violate uncertainty principles between position and momentum. This opens a remarkable possibility - could the many other radical effects of CTCs stand independent from the breaking of causality?

Here, we demonstrate that OTCs are remarkably powerful, and can replicate many defining operational benefits of CTCs. In sending a particle back in time - even when it interacts with nothing in the past - we can clone arbitrary quantum states to any fixed accuracy, and thus violate any uncertainty principle. Meanwhile, they also grant quantum processors additional computational power, allowing efficient solution of NP-complete problems. Our results hint that the remarkable power of Deustchian CTCs may survive the censorship of chronology protection. This drastically improves the potential of harnessing such power via alternative effects - such as certain models of gravitational time dilation~\cite{Pienaar13}. Thus, we open the possibility of testing the many radical protocols that harness CTCs in significantly less controversial settings.

\textbf{Framework}. In general relativity, causality can be violated due to the presence of spacetime wormholes that facilitate closed timelike curves (see Fig. \ref{Fig:CTC}). This allows a physical system $A$ to travel into its own causal past, and interact with its past self via some unitary $U$. The Deutschian model resolves potential paradoxes by enforcing temporal self-consistency~\cite{deutsch1991quantum, ralph2010information}, i.e.,
\begin{equation}\label{eq:Deutsch1}
  \rho_{\mathrm{CTC}} = \mathrm{Tr}_{\neq A}\left[U(\rhoin\otimes\rho_{\mathrm{CTC}})U^\dag\right],
\end{equation}
where $\rhoin$ denotes the initial state of the system, $\rhoctc$ the state it evolves to at the point of wormhole traversal and $\mathrm{Tr}_{\neq A}$ represents tracing over all systems other than $A$. Given a solution for $\rho_{\mathrm{CTC}}$, the final output of the process is given by
\begin{equation}\label{eq:Deutsch2}
\rhoout = \mathrm{Tr}_A\left[U(\rhoin\otimes\rho_{\mathrm{CTC}})U^\dag\right].
\end{equation}The many radical effects of CTCs rely on using specific self-interactions $U$ to break causality in different ways~\cite{Brun09,brun2013quantum,brun2003computers,Bacon04,aaronson2009closed}.

\begin{figure}[t]
\centering
\resizebox{8cm}{!}{\includegraphics[scale=1]{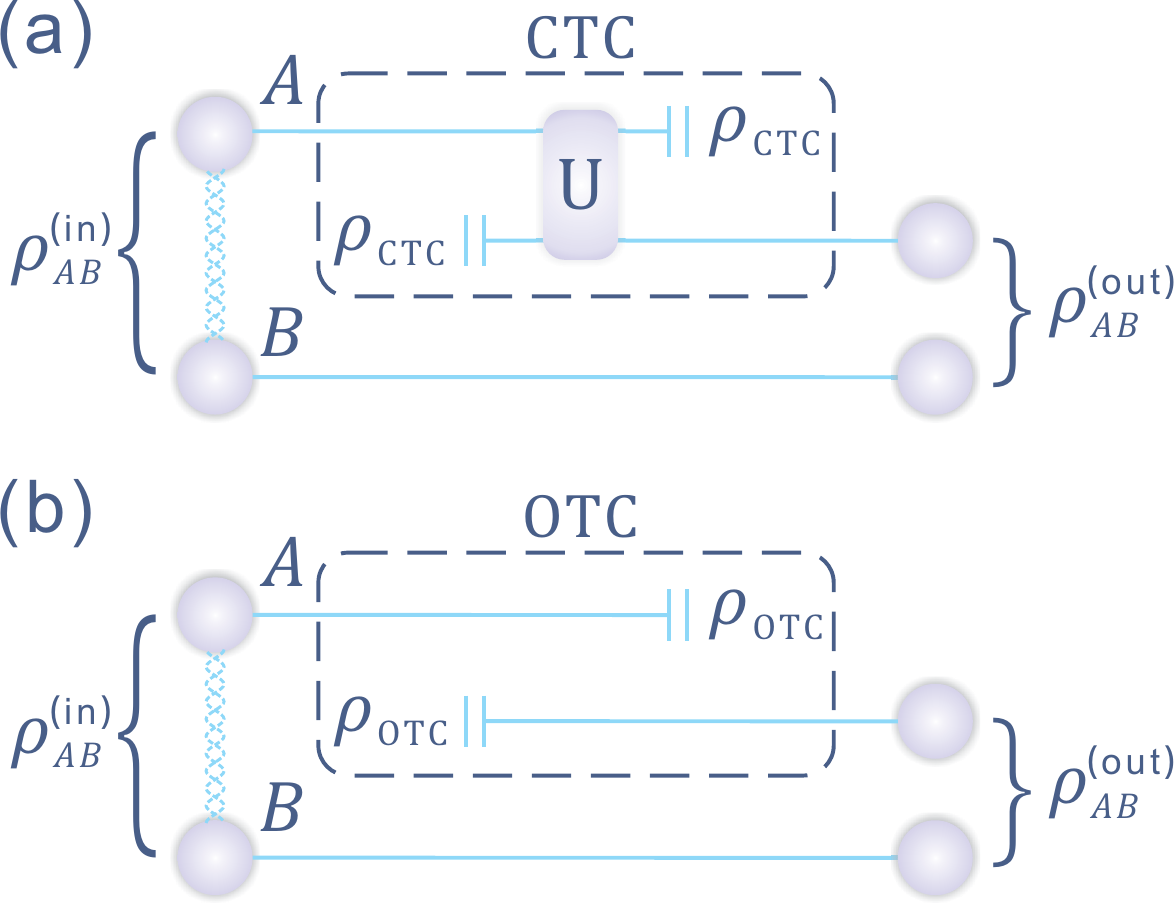}}
%\resizebox{8cm}{!}{\includegraphics[scale=1]{DOTC3.eps}}
  \caption{\textbf{CTCs and OTCs in presence of ancilla}. $A$ represents the system to be sent through the space-time wormhole, and $B$ some chronology respecting  system initially correlated with $A$. (a) In general CTCs, temporal self-consistency demands that $\rho_{\mathrm{CTC}}$ satisfies $\rho_{\mathrm{CTC}} = \mathrm{Tr}_{\neq A}[U(\rhoin\otimes\rho_{\mathrm{CTC}})U^\dag]$. (b) In the case of OTCs, this implies that system $A$ has state $\rho_{\mathrm{OTC}}= \mathrm{Tr}_{\neq A}\left[\rho_A\otimes\rho_{\mathrm{OTC}}\right]=\rho_{A}$ after interacting with its future self.
  }\label{Fig:OTC}
\end{figure}

Note that while the above analysis does not assume $\rhoin$ is pure, it \emph{only} applies for mixed inputs if $\rhoin$ represents one partition of a larger composite system that is pure. In the scenario where an input $\ket{\phi_k}$ is prepared with probability $k$, the dynamics of the CTC on each $\ket{\phi_k}$ must be analyzed separately. This is due to non-linearity, which implies differing unravellings of the density operator yield different outputs.

In OTCs, causality is preserved. The unitary $U$ is the identity - such that the time-travelling system does not interact with its causal past. Any observer in the frame of reference of $A$ can assign a valid chronology to all the events they witness. Meanwhile, to any outside observer, all events involving interactions with $A$ will respect causality. From an operational standpoint, there is no breaking of causality. If all information were classical, this entire procedure would only have the effect of desynchronizing $A$'s clock with that of an observer $B$.

Non-trivial effects, however, emerge when we consider quantum ancilla. Suppose we have access to a bipartite system $AB$ in state $\rho_{AB}$, where only one bipartition is sent through the OTC (see Fig.~\ref{Fig:OTC}b). The self-consistency relations imply
\begin{equation}\label{eq:decoherencor}
  \rhoout_{AB} = \mathrm{Tr}_{\neq A}[\rhoin_{AB}] \otimes \mathrm{Tr}_{A} [\rhoin_{AB}] = \rho_A \otimes \rho_B
\end{equation}
Thus, the OTC acts as a \emph{universal decorrelator} on $A$ - in sending a system $A$ though an OTC, we erase all quantum correlations between $A$ and the rest of the universe (and in particular, $B$). The resulting state, $\rho_A \otimes \rho_B$ fields identical local statistics with respect to the input $\rho_{AB}$, but none of its bipartite correlations. While this operation appears similar to trivial decoherence, it is non-linear, and shown to be impossible to synthesize with standard quantum dynamics~\cite{Terno99}.

This effect is associated with the monogamy of entanglement~\cite{ralph2010information} - a particle and its past self cannot be simultaneous entangled with the same external ancilla. While OTCs produce non-trivial dynamics when the input appears completely classical (e.g., when $\rhoin_{AB} = (\ket{00}\bra{00} + \ket{11}\bra{11})/2$), it applies only for mixed inputs if this mixedness is due to entanglement with some other system $C$. If we input $\ket{00}$ and $\ket{11}$ with equiprobability, then the dynamics of each input must be analyzed separately, and the OTC will have no effect.

\textbf{OTC enhanced measurement.} We first introduce \emph{OTC enhanced measurement}, a procedure that harnesses OTCs to measure an arbitrary observable $\hat{O}$ to any fixed precision. Specifically, given an unknown qudit ($d$ dimensional quantum system) in state $\rho$, we can determine $\langle \hat{O}\rangle = \mathrm{Tr}[\hat{O}\rho]$ to any desired accuracy $\delta > 0$ with negligible failure probability. This protocol functions as a building block for more sophisticated applications of open time-like curves, such as the solution of NP-complete problems and cloning of unknown quantum states.

\begin{figure}[t]
\centering
\resizebox{8cm}{!}{\includegraphics[scale=1]{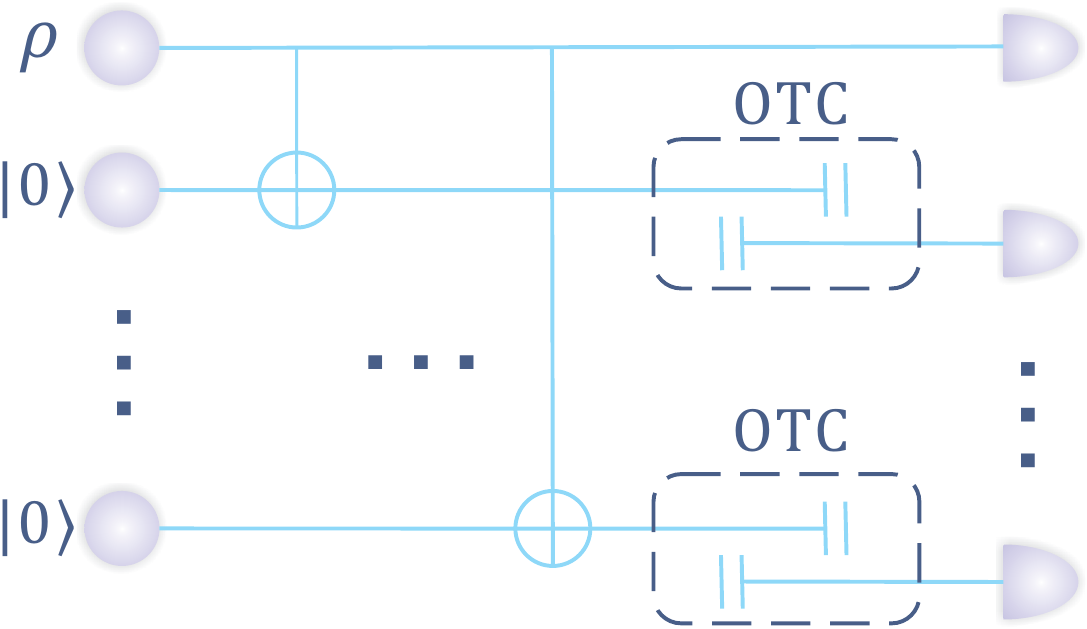}}
%\resizebox{8cm}{!}{\includegraphics[scale=1]{measure3.eps}}
  \caption{\textbf{Quantum circuit of OTC enhanced measurement}. The protocol first introduces $N$ ancilla qudits, all of which are initialized in state $\rho_E = \ket{0}\bra{0}$, where $\ket{0}$ is an eigenstate of $\hat{O}$. A sequence of $C_+$ gates then perfectly correlates each ancilla with $\rho$ with respect to $\hat{O}$ basis. The erasure of these correlations by OTCs, followed by $\hat{O}$ measurements on each individual qudit, allows determination of $\mathrm{Tr}[\hat{O}\rho]$ to a stand error that scales inversely with $N^2$.
  }\label{Fig:measure}
\end{figure}

The protocol is illustrated in Fig.~\ref{Fig:measure}. Let $\ket{j}: j = 0,1,\dots,d-1$ denote a basis that diagonalizes $\hat{O}$. On this basis, we introduce the two qudit controlled addition operator, $C_+\ket{i}\ket{j} = \ket{i}\ket{j+i}$, where addition is done modulo $d$. We then
\begin{enumerate}
  \item Prepare $N$ identical ancillary states in an eigenstates of $\hat{O}$, say $\ket{0}$.
  \item Apply the $C_+$ operations $N$ times, each controlled on $\rho$ and targeting a fresh ancilla state. This correlates $\rho$ with each of the $N$ ancillaries.
  \item Pass each of the ancillaries through an OTC to destroy all correlations in this $N+1$-partite system.
\end{enumerate}

This results in $N+1$ uncorrelated qudits, each in state $\rho_{\mathrm{diag}} = \sum_{i = 1}^d\rho_{ii}\ket{i}\bra{i}$, where $\rho_{ii}$ are the diagonal elements of $\rho$ in the $\hat{O}$ basis (see methods for details). Thus, each qudit exhibits identical statistics to $\rho$ when measured in the $O$ basis. In taking the mean of these measurements, we obtain an estimate for $\langle \hat{O}\rangle$. By the central limit theorem, the error of our estimate scales linearly with $1/\sqrt{N}$. In particular, provided the eigenvalues of $O$ are bounded, Hoeffding's bound implies we can estimate $\hat{O}$ to any desired accuracy $\delta$ and error rate $\epsilon$ using $O[1/\delta^2\log(1/\epsilon)]$ OTCs (see methods for details).

%Thus, we gain progressively better estimates of $\langle \hat{O}\rangle$ which each $N$. Application of the central theorem implies the error of our estimate will scale inversely with $N^2$.
%We now examine this protocol in more detail. By decomposing $\hat{O}$ and $\rho$ in the eigenbasis of $\hat{O}$ as $\hat{O} = \sum_{i = 1}^d o_i\ket{i}\bra{i}$ and $\rho = \sum_{i,j=1}^d\rho_{ij}\ket{i}\bra{j}$, one can easily see that measuring $\hat{O}$ yields $\sum_{i = 1}^d o_i\rho_{ii}$, which only depends on the diagonal terms of $\rho$. We can thus measure $\hat{O}$ on $\rho_{\mathrm{diag}}$ instead, where $\rho_{\mathrm{diag}} = \sum_{i = 1}^d\rho_{ii}\ket{i}\bra{i}$ is the density matrix containing only the diagonal components as $\rho$ \cite{Baumgratz13}. Upon the action of the decorrelators on the input states, it turns out that all the outputs are equal to $\rho_{\mathrm{diag}}$ (see Methods). For any observable $\hat{O}$ whose outputs belong to some finite range: $[o_{max}$ $o_{min}]$, this protocol allows us to estimate $\langle \hat{O}\rangle$ to within an arbitrary fixed precision $\delta$ and desired probability of error $\epsilon$ using order $O(d(o_{max} - o_{min})^2/\delta^2\log(1/\epsilon))$ OTCs.

\textbf{Solving NP-complete problems.} We take inspiration from Bacon~\cite{Bacon04}, who devised an efficient algorithm to solve the boolean satisfaction problem - a known NP-complete problem - using CTCs. We modify this algorithm to preserve causality - without losing efficiency. In the causality breaking algorithm, the key role of CTCs is to implement the non-linear map $S$ that maps an input qubit in state $\rho(n_z)$ to an output state $\rho(n_z^2)$, where $\rho(n_z) =  \frac{1}{2}\left(I + n_z \sigma_z\right)$ and $\sigma_z = \ket{0}\bra{0} -  \ket{1}\bra{1}$ denotes the Pauli $Z$ matrix (see methods for details).

\begin{figure}[t]
\centering
\resizebox{8cm}{!}{\includegraphics[scale=1]{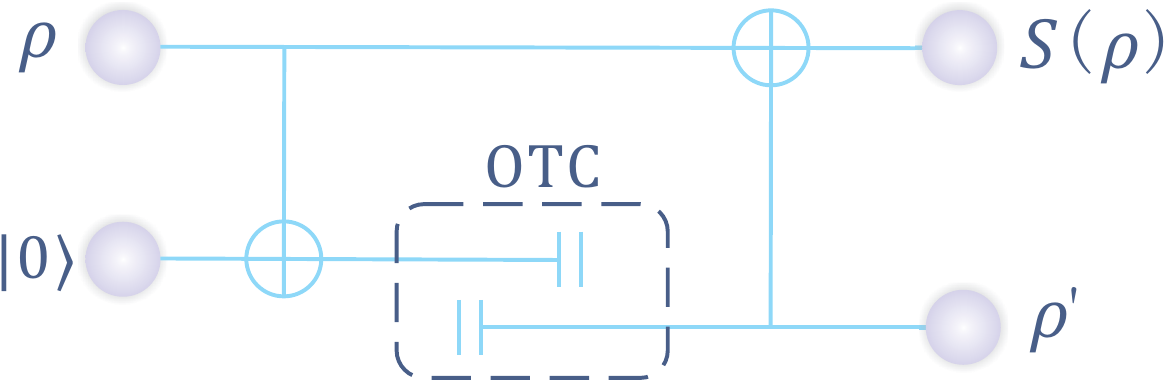}}
%\resizebox{8cm}{!}{\includegraphics[scale=1]{NP4.eps}}
  \caption{\textbf{Solving NP-complete problems with OTCs.} The key non-linear gate $S$, that takes $\rho(n_z)$ to $\rho(n_z^2)$, can be implemented by open timelike curves. This is achieved by use of a single OTC, applied between two successive $C_+$ gates.}\label{Fig:NP}
\end{figure}

This non-linear map can be replicated without breaking causality (See Fig.~\ref{Fig:NP}). Consider a special case of OTC enhanced measurement, with $\sigma_z$ as the observable of interest and a single ancilla. Consider an input qubit $\rho$ with matrix elements $\rho_{ij}$. Application of the OTC outputs two uncorrelated qubits, each in state $\rho_{\mathrm{diag}} = \rho_{00} \ket{0}\bra{0} + \rho_{11} \ket{1}\bra{1}$. Instead of measuring each in $\sigma_z$ directly, we apply a further $C_+$ gate controlled on the ancilla. After discarding the ancilla, the input qubit is now transformed to $S(\rho)$ as required.

In generating $S(\rho)$ using only OTCs, we can translate Bacon's algorithm into one that does not break causality. We note that as each call of $S(\rho)$ takes one OTC, the translation from CTCs to OTCs incurs no overhead on the number of times a particle needs to be sent through a spacetime wormhole. Thus, for the purposes of solving NP-complete problems, an OTC, together with one bit of entanglement, is at least as a powerful as a CTC.

\textbf{Cloning with OTCs.} Given an unknown input $\rho$, OTCs allow us to generated an unlimited number of clones to arbitrary fidelity. Our approach harnesses OTC enhanced measurements as a subroutine, which allows us to accurately determine $\mathrm{Tr}[\hat{M_i} \rho]$, for any observable $M_i$. First observe that this remains possible even if we are supplied with
\begin{equation}\label{eq:cloner}
   \rho' = s\rho + \frac{1-s}{d}I,
\end{equation}
a very noisy version of $\rho$. Here $I$ is the $d$-dimensional identity matrix, and $s$ is some fixed parameter such that $0 < s < 1$.

\begin{figure}[t]
\centering
%\resizebox{8cm}{!}{\includegraphics[scale=1]{Draw4.eps}}
\resizebox{8cm}{!}{\includegraphics[scale=1]{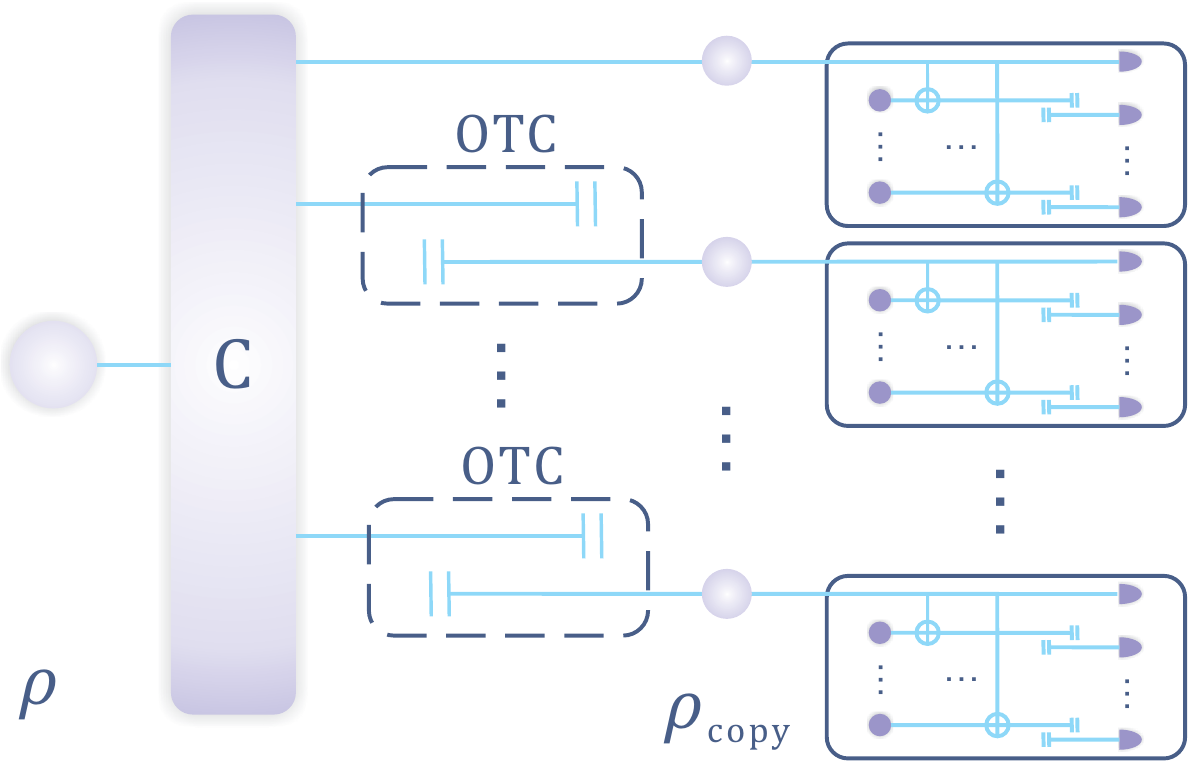}}
  \caption{\textbf{OTC Assisted Cloning}  An arbitrary qudit $\rho_S$ can be cloned to any desired fidelity. The process involves (i) application of a standard quantum cloner $C$ to generate $O(d^2)$ imperfect copies, and (ii) use of OTC enhanced measurements to measure different observables $M_i$ on each imperfect copy. We can choose $M_i$ to be informationally complete, and OTCs ensure that we can determine $\mathrm{Tr}{M_i \rho}$ to any desired precision. Thus this protocol can yield (to any fixed precision) the classical description of $\rho_S$.}\label{Fig:scheme}
\end{figure}

This observation, together with imperfect quantum cloners, form the basis of OTC enhanced cloning (Fig.~\ref{Fig:scheme}). In conventional quantum theory, an unknown quantum state $\rho$ can be cloned if we are given sufficiently many copies to perform accurate tomography~\cite{qubittomo_2002}.  One way to do this, is to use a set of $O(d^2)$ informationally complete measurements $\{M_i\}$, whose expectation values $\mathrm{Tr}(M_i \rho)$ has a one to one correspondence with the classical matrix description of $\rho$. Given only a single copy of $\rho$, this option is no longer valid.

OTC enhancement measurements alleviates this problem. We use standard methods to construct $O(d^2)$ imperfect clones in the form of Eq. \ref{eq:cloner}, where $s$ scales as $1/d$ for an optimal cloner \cite{Werner98}. Each clone is passed through an OTC to remove all entanglement between clones. An OTC enhanced measurement is then performed on each clone with respect to a different $M_i$. The outcomes of these measurements determine the density matrix of $\rho$. In methods, we show that by using $O(d^4/\delta_c^2\log{1/\epsilon_c})$ OTCs, we can ensure that each $\langle M_i\rangle$ is obtained to an accuracy of $\delta_c$ with fail probability $\epsilon_c$.
\

\textbf{Simple Example.} We illustrate these ideas by cloning a qubit. Here, the Pauli operators $\sigma_k, k = x,y,z$ is informationally complete - any $\rho$ is uniquely defined by the expectation values $n_k = \mathrm{Tr}[\sigma_k \rho]$. To determine each $n_k$, we first apply a universal 1-to-3 quantum cloner to obtain three imperfect clone of $\rho$, each in state $\rho' = (I + s\vec{n}\cdot\vec{\sigma})/2$ with $s = 5/9$ \cite{Scarani05}. These copies can be made independent via OTCs.

An OTC enhanced measurement of $\sigma_z$ is then performed on one such clone. We initialize $N$ ancilla qubits in state $\ket{0}$, and apply a CNOT gate between each ancilla and the clone (with the clone as the control qubit). In erasing the resulting correlations by sending each clone through an OTC, we obtain $N+1$ qubits, each in the state $(I + s n_z \sigma_z)/2$. Provided $N$ is sufficiently large, measurement of these qubits allows $n_z$ to be determined any desired accuracy with negligible error. Repetition of this process with $\sigma_x$ and $\sigma_y$ on the two remaining imperfect clones then yields complete information about $\rho$.

\textbf{Discussion.} Here, we demonstrated that the many defining operational benefits of closed timelike curves can be retained - without sacrificing causality. Our approach was to consider open timelike curves, where physical systems may travel back in time, but do not interact with themselves, or any other system in their past light-cone. In harnessing these open timelike curves, we developed methods to efficiently solve NP-complete problems, and cloning unknown quantum states to arbitrary fidelity. Many other radical effects attributed to CTCs come as natural consequences, such as distinguishing non-orthogonal states and violation of the Holevo limit. %While we have limited our discussion to discrete variables, many of the ideas generalize naturally to continuous variables. The use of quantum non-demolition interactions in place of $C_+$ gates for example, would allow OTC enhanced measurements of momentum or position.

This result highlights the intricate interplay between quantum theory and general relativity. If all physical systems have a well defined local reality, open timelike curves would have no operational effect. Our protocol thus fields no classical explanation. In the quantum regime, however, entanglement exists. While the local properties of a physical system are unaffected by open time like curves, their correlations with other chronology respecting systems are complete erased. In each of our protocols, this effect played a central role, allow us to replicate the many defining benefits of close timelike curves. This remarkable success propels us to conjecture whether entanglement assisted open timelikes curves are operationally equivalent to their causality breaking counterparts. Could one, for example, derive a map that takes any quantum circuit with CTCs, and engineer it in a way that does not break causality?

Preserving causality has significant benefits. Breaking causality is likely to be non-trivial, and opportunities to do so are negligible in the foreseeable future. This makes it unlikely for us to directly test the predictions of Deutschian CTCs. The preservation of causality in OTCs, however, suggest that its non-linear effects may be synthesized using alternative means. For instance, from the perspective of a chronology respecting observer, a particle sent through an OTC exhibits nothing more than time delay. Thus, in order to reconcile quantum field theory with non-hyperbolic space-times, gravitational time-dilation has been conjectured to share similar operational effects as OTCs \cite{Ralph09}. If true, our protocols suggest the exotic benefits of quantum processing in the general relativistic regime can be tested much sooner than previously expected.

\section{Methods}

\textbf{Scaling Analysis}. Execution of the OTC enhanced measurement with $N$ ancilla (and therefore $N$ uses of the OTC) to estimate $\langle \hat{O}\rangle$ gives an output $O_{\mathrm{est}} = \sum_k O_k/(N+1)$. We define the measurement as being successful if the estimate achieves a desired accuracy of $\delta$ (i.e., $|O_{\mathrm{est}} - \hat{O}| < \delta$). Application of Hoeffding's inequality \cite{hoeffding1963probability} gives failure probability $p_{f}$ that obeys
\begin{equation}\label{}
p_{f} \leq 2\exp\left[\frac{-2(N+1)\delta^2}{(O_\mathrm{max} - O_\mathrm{min})^2}\right].
\end{equation}
Here, $O_\mathrm{max}$ and $O_\mathrm{min}$ are the respective maximum and minimum eigenvalues of $\hat{O}$. Therefore
\begin{equation}\label{}
  N > \frac{(O_\mathrm{max} - O_\mathrm{min})^2}{2\delta^2}\log\frac{2}{\epsilon},
\end{equation}
OTC applications ensures a failure probability of no more than $\epsilon$. Provided $\hat{O}$ is bounded, this scales as $O[1/\delta^2 \log(1/\epsilon)]$.
 %Fig. \ref{Fig:measure}.

In OTC assisted cloning, we need to make $d^2$ informationally complete measurements, each to a desired accuracy $\delta > 0$ with negligible failure probability $\epsilon > 0$. Recall this is achieved via a $1 \rightarrow d^2$ universal cloner, whose imperfect copies are to be decorrelated via the use of OTCs (Fig. \ref{Fig:scheme}). For each of the $d^2$ copies, we apply an OTC enhanced measurement. To ensure this measurement is within accuracy $\delta$, an extra $O(d^2)$ overhead is required to compensate for the noise within the imperfect copies. The total number of OTCs required is then of order
\begin{equation}\label{}
  N > O\left[d^4\left(\frac{1}{2\delta^2}\log\frac{2}{\epsilon}\right)\right],
\end{equation}
where $O_\mathrm{max} - O_\mathrm{min}$ is equal to 1 for members of the information complete basis.

\textbf{Solving NP-complete problems.} Here we outline explicitly how a non-linear map that takes $\rho(n_z)$ to $\rho(n_z^2)$ allows the efficient solution of NP-complete problems. Specifically we study the satisfaction problem: \emph{Given a Boolean function $f:\{0,1\}^n\rightarrow\{0,1\}$, specified in conjunctive normal form, does there exist a satisfying assignment $(\exists x|f(x)=1)$?} This problem is known to be NP-complete.

Bacon~\cite{Bacon04} showed that this problem can be efficiently solved if the following gate is available,
\begin{equation}\label{}
  S(\rho) = \frac{1}{2}\left(I + n_z^2\sigma_z\right),
\end{equation}
where the input qubit is $\rho = (I + \vec{n}\cdot\vec{\sigma})/2$, with $\vec{n}$ being the Bloch sphere vector and $\vec{\sigma}$ is a 3 component vector, whose entries are the Pauli matrices.

In Fig. \ref{Fig:NP} we demonstrated how this gate can be synthesized using OTCs. This established, the satisfaction problem is efficiently solved as follows:
\begin{enumerate}
  \item Prepare $n$ ancillary qubits in the state $1/\sqrt{2^n}\sum_{i=0}^{2^n-1}\ket{i}$ and a target qubit in state $\ket{0}$.
  \item Act the unitary
      \begin{equation}\label{}
        U_f = \sum_{i=0}^{2^n-1}\ket{i}\bra{i}\otimes\sigma_x^{f(i)},
      \end{equation}
      on this system (with the last qubit representing the target). Tracing out the ancillary qubits leaves the target in
      \begin{equation}\label{}
        \rho = \frac{1}{2}\left[1+\left(1-\frac{s}{2^{n-1}}\right)\sigma_z\right],
      \end{equation}
      where $s$ is the number of $x$ satisfying $f(x)=1$.
  \item  Apply $S$ to the target via the use of OTCs (See Fig. \ref{Fig:NP}). Repeat this step $p$ times to get
    \begin{equation}\label{}
        \rho_p = \frac{1}{2}\left[1+\left(1-\frac{s}{2^{n-1}}\right)^{2^p}\sigma_z\right],
      \end{equation}

    %\item Run step (1)-(3) $q$ times.
\end{enumerate}

%\textcolor{blue}{Sorry but I am struggling to understand this paragraph. Can we be clearer about p and q, and also clarify what we meant in the sentence 'By measuring the $\sigma_z$ basis, whether there is output of $-1$ will tell us whether $s \neq 0$ or $s=0$'. I assume the proabbility of failure will make sense to me after we fix this sentence.}
Notice that, we could easily check the case of  $s=2^n$. Thus we only need to distinguish between $s=0$ and $0<s<2^n$.
With the limit of $p\rightarrow\infty$, the two output states corresponding to the cases of $s=0$ and $0<s<2^n$ are $\rho_p= \ket{0}\bra{0}$ and $\rho_p \rightarrow I/2$, respectively. By performing measurement in the $\sigma_z$ basis, one can distinguish the two types of output states $\ket{0}\bra{0}$ and $I/2$, that is, the case of  $s=0$ and $0<s<2^n$, with failure probability being $1/2$. By repeating these steps more times, say, $q$, the failing probability exponentially decays. For finite $p$ and $q$ that are polynomial in $n$, the probability of failure is given by  \cite{Bacon04}
\begin{equation}\label{}
  P_{fail}=\frac{1}{2^q}\left[1+\left(1-\frac{s}{2^{n-1}}\right)^{2^p}\right]^q.
\end{equation}

\textbf{Acknowledgements.} We thank D. Terno for helpful discussions. The
research is supported by the National Research Foundation and Ministry of Education in Singapore, the Tier 3  MOE2012-T3-1-009 Grant ``Random numbers from quantum processes", the Australian Research Council Centre of Excellence for Quantum Computation and Communication Technology Project number CE110001027, the National Basic Research Program
of China Grant 2011CBA00300, 2011CBA00302 and the National Natural
Science Foundation of China Grant 11450110058, 61033001, 61361136003.

%%%%%%%%%%%%%%%%%%%%%%%%%%%%%%%%%%%%%%%%
% choose a style
%\bibliographystyle{ieeetr}
%\bibliographystyle{unsrt}
\bibliographystyle{apsrev4-1}
%%%%%%%%%%%%%%%%%%%%%%%%%%%%%%%%%%%%%%%%

%%%%%%%%%%%%%%%%%%%%%%%%%%%%%%%%%%%%%%%%
% choose a .bib file
%\bibliography{BibOTC}
%%%%%%%%%%%%%%%%%%%%%%%%%%%%%%%%%%%%%%%%

%

\section{Supplementary material}

\textbf{Details of OTC Enhanced measurement}. The protocol in Fig.~\ref{Fig:measure} computes the expectation value $\langle \hat{O}\rangle = {\rm Tr}[\rho \hat{O}]$ to any desired accuracy, when supplied with an unknown quantum state $\rho$ and $N$ ancillary states each initialized in an eigenstate $\ket{0}$, of $\hat{O}$. First $N$ controlled addition operations, $C_+: \ket{i,j} \rightarrow \ket{i,i+j}$, are applied to produce a GHZ-like input state,
\begin{equation}\label{eq:correlatedoutput}
  \rho_{\mathrm{in}}=\sum_{i,j=1}^d\rho_{ij}\ket{ii\dots i}\bra{jj\dots j}.
\end{equation}
If we tried to directly  measure a qudit in this state, then  we would get outcome $i$ with probability $\rho_{ii}$, and simultaneously collapse the other qudits onto $\ket{i\dots i}\bra{i\dots i}$. Therefore subsequent measurements will not provide any extra statistics on $\langle \hat{O}\rangle$. To resolve this problem, we use OTCs to erase the correlations in this GHZ-like state $N$ times, resulting in the output
\begin{equation}\label{eq:ouput}
\begin{aligned}
 \rhoout= \left(\sum_{i=1}^d\rho_{ii}\ket{i}\bra{i}\right)^{\otimes N+1}  = \rho_{\mathrm{diag}}^{\otimes N+1}.
\end{aligned}
\end{equation}
This corresponds to $N+1$ uncorrelated qudits, each in the diagonal state $\rho_{\mathrm{diag}}$.

This protocol highlights the key role of OTCs, and why their role of erasing correlations is remarkably useful. They take the highly correlated state in Eq. (\ref{eq:correlatedoutput}) and erase all multipartite correlations, while preserving the local statistics of each individual qudit. This allows each qudit to be measured independently without disturbing the others, and thus information collected from each measurement is statistically independent.

\end{document}